\documentclass[12pt]{article}
\usepackage{epsfig}\parskip 5pt plus 1pt
\newcommand{\be}{\begin{equation}}
\newcommand{\ee}{\end{equation}}
\newcommand{\bea}{\begin{eqnarray}}
\newcommand{\eea}{\end{eqnarray}}
\newcommand{\ba}{\begin{eqnarray}}
\newcommand{\ea}{\end{eqnarray}}
\newcommand{\helio}{^6{\rm He}}
\newcommand{\neon}{^{18}{\rm Ne}}

\def\simge{\mathrel{%
   \rlap{\raise 0.511ex \hbox{$>$}}{\lower 0.511ex \hbox{$\sim$}}}}
\def\simle{\mathrel{
   \rlap{\raise 0.511ex \hbox{$<$}}{\lower 0.511ex \hbox{$\sim$}}}}

\begin{document}
\thispagestyle{empty}
\vspace*{1cm}
\begin{center}
{\Large{\bf High energy beta beams without massive detectors } }\\
\vspace{.5cm}
F.~Terranova$^{\rm a}$, A.~Marotta$^{\rm b}$, P. Migliozzi$^{\rm b}$,  
M.~Spinetti$^{\rm a}$ \\
\vspace*{0.5cm}
$^{\rm a}$ I.N.F.N., Laboratori Nazionali di Frascati,
Frascati (Rome), Italy \\
$^{\rm b}$ I.N.F.N., Sezione di Napoli, Naples, Italy 
\end{center}

\vspace{.3cm}
\begin{abstract}
\noindent
In this paper, the possibility to exploit a high energy beta beam
without massive detectors is discussed. The radioactive ions are
boosted up to very high $\gamma$ with the neutrino beam pointing
towards an instrumented surface located at a moderate baseline
(e.g. from CERN to the Gran Sasso Laboratories). $\nu_e \rightarrow
\nu_\mu$ oscillations and their CP conjugate are tagged as an excess
of horizontal muons produced in the rock and tracked by the low-mass
instrumented surface installed in one of the LNGS experimental
halls. We show that the performance of this complex for what concerns
the determination of the $\theta_{13}$ angle of the leptonic mixing
matrix is comparable with the current low-$\gamma$ design based on a
gigantic water Cherenkov at Frejus.
\end{abstract}

\vspace*{\stretch{2}}
\begin{flushleft}
  \vskip 2cm
{ PACS: 14.60.Pq, 14.60.Lm} 
\end{flushleft}

\newpage

\section{Introduction}
\label{introduction}

Over recent years there has been a marked growth of interest in the
development of non-conventional neutrino sources. An appealing
proposal has been put forward by P.~Zucchelli~\cite{Zucchelli:sa} in
2002 under the name of ``beta beam''. It is based on the production of
intense beams of $\beta$-unstable heavy ions. The ions are accelerated
to a given energy and stored in a decay ring with long straight
sections pointing towards a far detector. Their decays produce a pure
and intense $\nu_e$ ($\bar{\nu_e}$) beam whose spectrum depends solely
on the $\beta$-decay kinematics. The advantages of this configuration,
the possibility to explore subdominant $\nu_e \rightarrow \nu_\mu$
oscillations at the atmospheric scale with unprecedented sensitivity
and, hence, to extract the $\theta_{13}$ and $\delta$ parameters of
the leptonic mixing matrix (PMNS~\cite{PMNS}) have been discussed by
several
authors~\cite{Mezzetto:2003ub,Bouchez:2003fy,Burguet-Castell:2003vv}. In
particular, it has been noted than an European beta beam complex could
leverage existing facilities at CERN and complement the EURISOL
physics program~\cite{betabeams_moriond}. The latter foresees the
construction of an intense proton driver for a new generation of
radioactive beams. In fact, EURISOL is a significant extension of the
program presently being carried out using the first-generation
radioactive ion beam facilities in nuclear physics and nuclear
astrophysics. It is aimed at increasing the variety of exotic ions
produced and their yields by orders of magnitude beyond those
presently available. Hence, a CERN-based beta beam complex would
exploit the EURISOL ion source and the CERN PS/SPS acceleration
complex.  Only the dedicated hippodrome-like decay ring should be
built on purpose. In its most popular configuration $\nu_e$ are
produced by $\neon$ ions and $\bar{\nu}_e$ by
$\helio$~\cite{Bouchez:2003fy}; the ions are accelerated by the SPS up
to $\gamma \sim 100$ ($\neon$) and $\gamma \sim 60$ ($\helio$). The
ratio between the two boost factors is fixed by the equalization of
rigidity, i.e. the need of accumulating simultaneously into the same
ring ions with different $Z$. The corresponding neutrinos emerge with
energies below 1 GeV. In order to observe neutrino oscillations at the
peak of the oscillation probability the detector must be located at
$\sim$130~km from the source, matching, for instance, the distance
from CERN to Frejus. It has already been
noted~\cite{Burguet-Castell:2003vv} that low-$\gamma$ choice is, in
principle, quite unfortunate. A low-$\gamma$ beta beam aimed at the
observation of $\nu_\mu \rightarrow \nu_e$ oscillations at the
atmospheric scale needs a gigantic detector located at $L_1\simeq
130$~km as the proposed 1~Mton water Cherenkov at
Frejus~\cite{mton_at_frejus}. The size of the detector must overcome
the smallness of the cross section at mean $\nu$ ($\bar{\nu}$)
energies of the order of 0.3~(0.2)~GeV.  A higher energy beta beam and
a detector located at a farther location $L_2$, tuned to operate at
the peak of the oscillation probability, would provide a flux similar
to the low-$\gamma$ option since the neutrino fluxes increase
quadratically with the boost factor and decrease as $L^2$. However,
operating at larger $\gamma$ show up additional advantages due to the
enhanced $\nu_\mu$ CC cross section, which depends linearly on the
neutrino energy.  Hence, as a first approximation, we expect the
sensitivity to the subdominant $\nu_e \rightarrow \nu_\mu$ channel to
grow as $\gamma_{L_2}/\gamma_{L_1}$, i.e. as the ratio of the boost
factors needed to be at the peak of the oscillation probability for
the distance $L_1$ and $L_2$, respectively. A further increase of
$\gamma$ with respect to $\gamma_{L_2}$ would cause a further
quadratic rise of the flux compensated by a quadratic drop of the
oscillation probability (``off-peak''
configuration~\cite{Migliozzi:2003pw}). If the dependence of the
oscillation probability on the PMNS parameters were the same in the
``on-peak'' and ``off-peak'' configuration, this would, anyhow, imply
an increase of sensitivity due to the further growth of the cross
section.  Therefore, even under this condition, the possibility to run
the beta beams in an off-peak configuration could be worth being
investigated.  However, this scenario turns out to be even more
attractive if we consider the detector technologies that could be
exploited to observe $\nu_e \rightarrow \nu_\mu$ oscillations at high
$\nu_\mu$ energies. In particular, the $\nu_e \rightarrow \nu_\mu$ and
$\bar{\nu}_e \rightarrow \bar{\nu}_\mu$ channels could be observed as
an excess of high energy ($>$~1-2~GeV) muons from the rock of an
underground laboratory tagged by an instrumented surface installed
into the cavern. Since the muon range in the rock grows linearly with
the muon energy, the effective mass of the rock that contributes to
the event rate adds a further linear dependence on $\gamma$ so that a
nearly quadratic increase of the sensitivity due to the higher beta
beam energy is gained.  Clearly, if the background can be kept under
control, this configuration allows an enormous simplification and
reduction of cost with respect to the Mton water Cherenkov option,
especially if the detector can be installed in pre-existing halls as
the ones of the Gran Sasso INFN Laboratories. In this paper, we
demonstrate the feasibility of this design and determine its
performance. The structure of the subdominant $\nu_e \rightarrow
\nu_\mu$ oscillations for off-peak scenarios and the functional
dependence of the sensitivity to the parameters of the PMNS on the
beta beam energy is derived in Sec.~\ref{sec:probosc}.  The detector
concept and the main backgrounds are discussed in
Sec.~\ref{sec:detector}. Its sensitivity to the (1-3) sector of the
PMNS is computed in Sec.~\ref{sec:sensitivity}.

\section{Oscillations at a high $\gamma$ beta beam}
\label{sec:probosc}

\begin{figure}
\centering
\epsfig{file=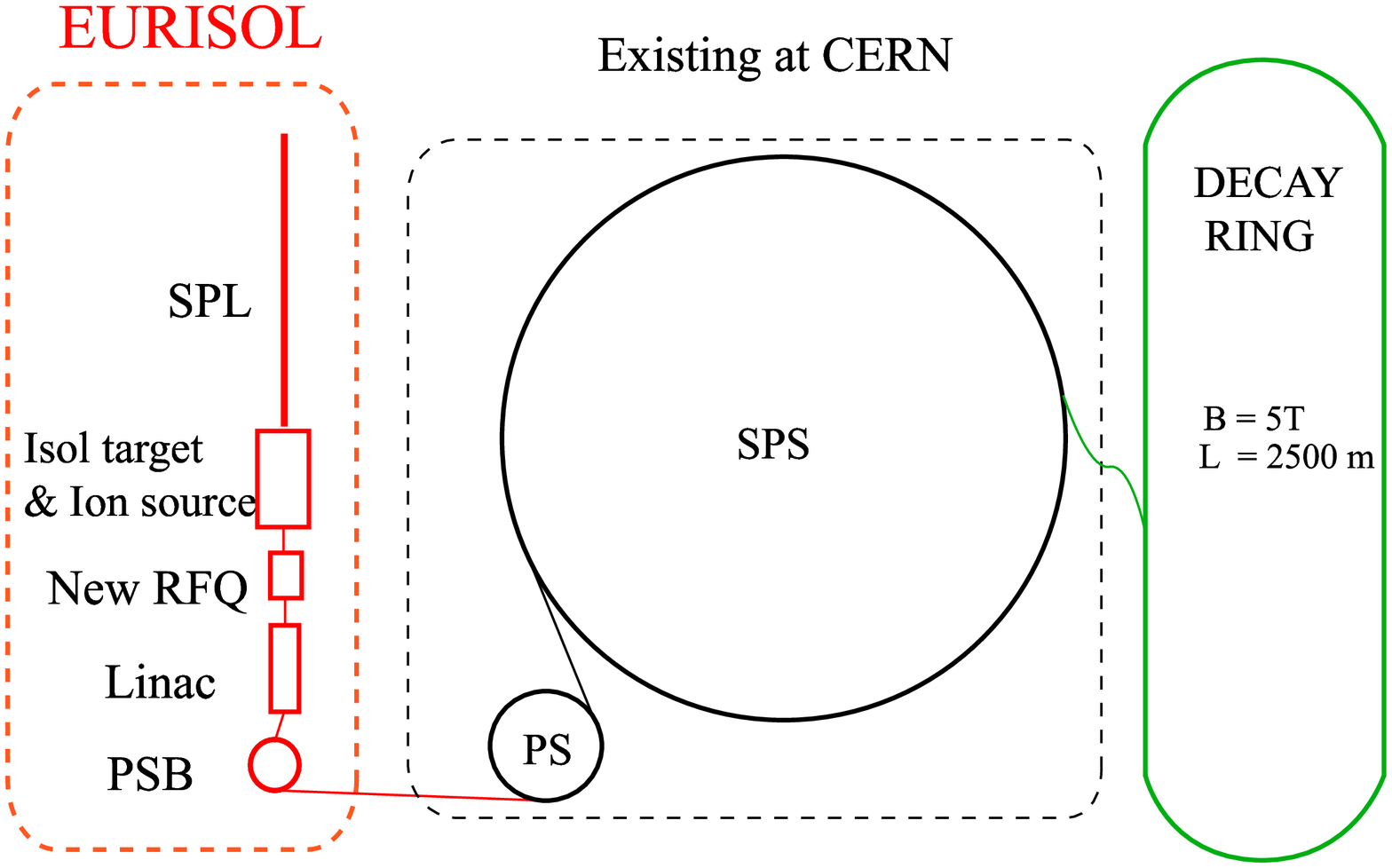,width=10cm}
\caption{The beta beam complex based on CERN facilities in the 
low-$\gamma$ configuration.}
\label{fig:complex}
\end{figure}

The acceleration of radioactive ions is a prominent technique for
nuclear physics studies and several facilities have been developed
worldwide. The use of short-lived $\beta$-decay isotopes transforms
these facilities into high intensity sources of pure $\nu_e$ or
$\bar{\nu}_e$. The source has practically no contamination from other
flavors and a well defined energy spectrum that depends on the
kinematic of $\beta$-decay. The annual flux for a far detector located
at a distance $L$ and aligned with the boost direction of the parent
ion is~\cite{Burguet-Castell:2003vv}:

\ba \left.{d\Phi \over dS dy}\right|_{\theta\simeq 0} \simeq
{N_\beta \over \pi L^2} {\gamma^2 \over g(y_e)} y^2 (1-y)
\sqrt{(1-y)^2 - y_e^2}, \ea
where $0 \leq y={E_\nu \over 2 \gamma E_0} \leq 1-y_e$, $y_e=m_e/E_0$
and 
\be 
g(y_e)\equiv {1\over 60} \left\{ \sqrt{1-y_e^2} (2-9 y_e^2 - 8
y_e^4) + 15 y_e^4 {\mathrm Log} \left[{y_e \over
1-\sqrt{1-y_e^2}}\right]\right\} 
\ee
In this formula $E_0$ represents the electron end-point energy, $m_e$
the electron mass, $E_\nu$ the energy of the final state neutrino and
$N_\beta$ is the total number of ion decays per year. A beta beam
facility based on existing CERN machines has been discussed
in~\cite{Bouchez:2003fy,betabeams_moriond,Autin:2002ms}. The protons
would be delivered by the Super Proton Linac (SPL)~\cite{SPL}. This
driver has been studied at CERN in the framework of the neutrino
factory~\cite{Apollonio:2002en} but could be an essential part of the
EURISOL complex. The SPL would provide 2.2~GeV (kinetic energy)
protons with an intensity of 2~mA. The targets for ion production
would be similar to the ones envisioned by EURISOL. In particular,
antineutrinos could be produced by $\helio$ decays (a $\beta^-$
emitter with $E_0-m_e=3506.7$~keV and a 806.7 ms half life) from a target
consisting either of a water cooled tungsten core or of a liquid lead
core which works as a proton to neutron converter surrounded by
beryllium oxide. Neutrinos could result from the $0^+ \rightarrow 1^+$
$\beta^+$ decay of $\neon$ ($E_0-m_e=3423.7$~keV and half life of
1.672~s); the isotope can be produced by spallation reactions and, in
this case, protons directly hit a magnesium oxide target. The ions can
be further accelerated using the EURISOL linac and the PS/SPS complex
(see Fig.~\ref{fig:complex}) and sent to the decay ring.  In this case
the nominal $\gamma$ is fixed by the present SPS design.  It
corresponds to $\gamma \sim 60$ for $\helio$ and $\gamma \sim 100$ for
$\neon$. Higher values of $\gamma$ can be achieved upgrading the SPS
with superconducting magnets or making use of the LHC. The top
rigidity available at the LHC could allow for $\gamma=2488$ ($\helio$)
and $\gamma=4158$ ($\neon$)~\cite{talk_lindroos} even if in this case
the construction of the decay ring would be challenging.  In
particular, due to the high rigidity, it would be unrealistic to build
a ring with a curved-over-straight section ratio similar to the
low-$\gamma$ option~\cite{Bouchez:2003fy}. Here, the ratio is
0.94~km/2.5~km (see Fig.~\ref{fig:complex}) and the useful fraction of
decays is limited by the decays occurring when the bunch is located in
the return straight section (i.e. the boost has opposite direction
w.r.t. the far detector). The overall live-time, i.e. the fraction of
decays occurring at the straight section pointing to Frejus, is
36\%. If the size of the straight section is kept unchanged (2.5~km)
in a high-$\gamma$ configuration, the live-time is limited by the size
of the curved parts, which in turns is related to the maximum magnetic
field achievable with superconducting, radiation-hard magnets. It
drops at the level of 20\% and 10\% for a straight section of 2.5~km
and a curved section comparable with the size of the SPS and the LHC,
respectively. The corresponding loss of statistics can be easily and
more cheaply recovered by instrumenting a larger surface at the far
location (e.g. more than one experimental hall at LNGS - see
below). The actual cost of the decay ring depends on the maximum field
available and on the possibility to use a significant fraction of the
accelerating ring also in the storage phase.  In the low-$\gamma$
design $2.9\times 10^{18}~\helio$ and $1.1\times 10^{18}~\neon$ decays
per year are expected.  If the LHC were used, some injection losses
would be expected due to the different optics; these losses could be
compensated by an increase of the number and of the length of the
bunches~\cite{Burguet-Castell:2003vv}.

The number of oscillated $\nu_e \rightarrow \nu_\mu $ events per year
that can be observed at a distance $L$ with a detector of $M$~kton is
\ba
N_{osc} & = & M \ 10^9 \ N_A N_\beta \int_0^{2\gamma(E_0-m_e)} dE 
\frac{\gamma^2}{\pi L^2 g(y_e)}   \frac{E^2(2\gamma E_0-E)}{(2\gamma E_0)^4}
\times \nonumber \\ 
& & \times \sqrt{ \left(1-\frac{E}{2\gamma E_0} \right)^2-y_e^2 } \ 
\sigma^{CC}_{\nu_\mu}(E) \  P(\nu_e \rightarrow \nu_\mu) \ \epsilon(E)
\label{equ:probosc}
\ea
where $N_A$ is the Avogadro's number, $\gamma$ is the boost factor of
the beta beam complex, $E$ is the neutrino energy, $\epsilon(E)$ 
the detector efficiency, $\sigma^{CC}_{\nu_\mu}(E)$ the $\nu_\mu$~CC
cross section at a given energy and $P(\nu_e \rightarrow \nu_\mu)$ the
oscillation probability.  The latter depends on the baseline $L$, the
energy $E$ and the parameters of the PMNS matrix. In particular, in the
energy/baseline range of interest, $P(\nu_e \rightarrow \nu_\mu)$ can
be expressed as~\cite{cervera_freund}:

\begin{eqnarray}
P(\nu_e \rightarrow \nu_\mu) & \simeq & \sin^2 2\theta_{13} \, \sin^2
\theta_{23} \frac{\sin^2[(1- \hat{A}){\Delta}]}{(1-\hat{A})^2}
\nonumber \\ & + & \alpha \sin 2\theta_{13} \, \xi \sin \delta
\sin({\Delta}) \frac{\sin(\hat{A}{\Delta})}{\hat{A}}
\frac{\sin[(1-\hat{A}){\Delta}]}{(1-\hat{A})} \nonumber \\ &+& \alpha
\sin 2\theta_{13} \, \xi \cos \delta \cos({\Delta})
\frac{\sin(\hat{A}{\Delta})}{\hat{A}} \frac{\sin[(1-\hat{A}){\Delta}]}
{(1-\hat{A})} \nonumber \\ &+& \alpha^2 \, \cos^2 \theta_{23} \sin^2
2\theta_{12} \frac{\sin^2(\hat{A}{\Delta})}{\hat{A}^2} \nonumber \\ &
\equiv & O_1 \ + \ O_2(\delta) \ + \ O_3(\delta) \ + \ O_4 \ \ .
\label{equ:probmatter}
\end{eqnarray}
In this formula $\Delta \equiv \Delta m_{31}^2 L/(4 E)$ and
the terms contributing to the Jarlskog invariant are split into the
small parameter $\sin 2\theta_{13}$, the ${\cal O}(1)$ term $\xi
\equiv \cos\theta_{13} \, \sin 2\theta_{12} \, \sin 2\theta_{23}$ and
the CP term $\sin \delta$; $\hat{A} \equiv 2 \sqrt{2} G_F n_e E/\Delta
m_{31}^2$ with $G_F$ the Fermi coupling constant and $n_e$ the
electron density in matter. Note that the sign of $\hat{A}$ depends on
the sign of $\Delta m_{31}^2$ which is positive (negative) for normal
(inverted) hierarchy of neutrino masses. In the following we assume
the present best fits for the solar and atmospheric parameters:
$\Delta m^2_{21}=7.3 \times 10^{-5}$~eV$^2$, $\sin^2
2\theta_{12}=0.8$, $|\Delta m^2_{31}|=2.5 \times 10^{-3}$~eV$^2$,
$\sin^2 2\theta_{23}=1$~\cite{Maltoni:2003da}.

\begin{figure}
\centering
\epsfig{file=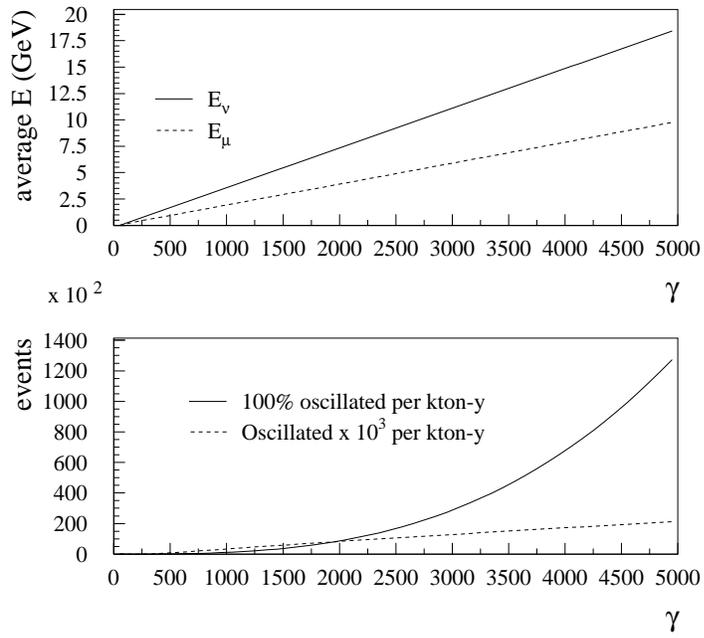,width=10cm}
\caption{(Upper plot) Average neutrino and muon energy versus $\gamma$
at a baseline of 732~km.  (Lower plot) The continuous line represents
the number of oscillated events per kton-y assuming 100\% conversion
probability. The dashed line shows the number of events per kton-y for
$\theta_{13}=3^\circ$, $\delta=0^\circ$ and normal neutrino hierarchy
($\Delta m_{13}>0$).  Rates are computed for assuming perfect detector
efficiency ($\epsilon(E)\equiv1$).}
\label{fig:func_behav}
\end{figure}

Far from the oscillation peak ($\Delta \ll 1$,  $|(1-\hat{A})\Delta| \ll 1$)
the functional dependence of $P(\nu_e \rightarrow \nu_\mu)$ becomes
similar to the one of CNGS~\cite{Migliozzi:2003pw}, i.e. 
\ba
P(\nu_e \rightarrow \nu_\mu) \simeq 
\Delta^2 \ \left[ \sin^2 2\theta_{13} \, \sin^2
\theta_{23} +  \alpha
\sin 2\theta_{13} \, \xi \cos \delta  \right]
\ea
and the oscillation probability mainly depends on $\theta_{13}$ and
$\cos \delta$.  Matter effects are strongly suppressed and CP
asymmetries appear only at the subleading order $O_2(\delta)$.
However, even in the highest $\gamma$ scenario, the mean neutrino
energy remains significantly lower than CNGS and the cancellation of
the $O_2(\delta)$ term is not complete. Hence, the $\bar{\nu}_e$ flux
(``antineutrino run'')\footnote{The antineutrino run is done in
parallel with the neutrino one, since it is possible to circulate
$\helio$ and $\neon$ ions simultaneously in the decay ring and exploit
the time structure of the beam to separate the two contributions at
the far detector.}  still contributes to constrain the
$(\theta_{13},\delta)$ parameter space. At fixed baseline $L$, an
increase of $\gamma$ implies a quadratic increase of the flux while an
increase of the mean neutrino energy $\langle E \rangle $ causes a
quadratic decrease of the oscillation probability, a linear increase
of the cross section ($\sigma^{CC}_{\nu_\mu} \sim E$ in the
region dominated by deep inelastic scattering)  and a linear increase
of the mean primary muon energy. Clearly, $\gamma$ and $\langle E
\rangle$ are fully correlated and $\langle E \rangle \sim \gamma$:
\begin{eqnarray}
\langle E \rangle & = &
  \frac{ \int_0^{2\gamma(E_0-m_e)} E \left[ \frac{d\Phi}{dE} \right] dE }
{  \int_0^{2\gamma(E_0-m_e)}  \left[ \frac{d\Phi}{dE} \right] dE }   \nonumber 
\\
& = & 2 \gamma E_0 \ \frac{ \int_{y_e}^1 (1-z)^3 z \sqrt{z^2-y_e^2} }
{ \int_{y_e}^1 (1-z)^2 z \sqrt{z^2-y_e^2} } \nonumber \\
& = & 2 \gamma E_0 \ f(y_e)
\end{eqnarray}
This functional behavior is depicted in
Fig.~\ref{fig:func_behav}. The upper plot shows the average neutrino
energy $\langle E \rangle$ versus $\gamma$. The average muon energy
for the $\nu_\mu N \rightarrow \mu^- X$ final state is also shown.
The lower plot represents the number of oscillated events per
kton-year as a function of $\gamma$ assuming 100\% conversion
probability (Eq.(\ref{equ:probosc}) with $P_{\nu_e \rightarrow
\nu_\mu}\equiv 1$) and an isoscalar target. The baseline is fixed at
$L=732$~km, i.e. at the location of the Gran Sasso Laboratories
(LNGS). The number of oscillated events ($\times 10^{3}$) per kton-y
for $\theta_{13}=3^\circ$, $\delta=0^\circ$ and normal neutrino
hierarchy ($\Delta m^2_{13}>0$) assuming perfect detector efficiency
is also shown.  Due to the assumption $\epsilon(E)=1$, the linear
increase of the average muon energy, resulting in a further linear
rise of the effective fiducial mass, is not exploited. This issue will
be discussed in the next section.

\section{The detector concept}
\label{sec:detector}

One of the most remarkable features of the beta beams is the clearness
of the final state to be observed. This advantage is less evident for
low-$\gamma$ configurations (neutrinos in the sub-GeV energy range);
here, the total cross section is dominated by quasi-elastic
interactions whose kinematic is obscured by Fermi motion. The latter
makes impossible to extract the full spectral information from the
final state reconstruction.  In this energy range, a massive water
Cherenkov performing a pure counting experiment could be an
appropriate detector.  At higher neutrino energies, the $\nu_\mu$
interactions are mainly deep inelastic and the use of denser detectors
becomes rather attractive. For a counting experiment, a dense tracking
detector provides a strong muon/pion and muon/electron separation and
identifies the neutrino direction through the reconstruction of the
muon track. Moreover, if the interaction vertex is contained, a
kinematic analysis is possible and greatly contributes to the
determination of the PMNS parameters~\cite{Burguet-Castell:2003vv}.
$\nu_\mu$~CC interactions occurring into the rock which surrounds the
detector can be exploited as well. In this case the rock acts as a
massive target. Electrons and pions mainly interact before reaching
the surface of the experimental hall and the tracking capability of
the detector can be used to veto punch-through hadrons and reconstruct
the residual muon energy at the entrance of the hall. The main
difference with respect to the case of fully instrumented volume is
that the kinematic analysis is deteriorated by the energy loss through
the rock but the event rate scales no more as the tracking {\it
volume} but it is proportional to the tracking {\it surface}. Hence,
an instrumented surface located in an existing deep underground hall
would imply order-of-magnitude cost reductions compared to a massive
water Cherenkov. In the following we consider an instrumented surface
(15$\times$15 m$^2$) installed in one of the halls of LNGS. The
tracking device is made of vertical iron walls interleaved with active
detectors. The granularity is chosen to guarantee an angular
resolution of a few degrees for horizontal muons. The overall
thickness of the iron must be appropriate to effectively separate
pions from muons at energies greater than 1-2~GeV.  These requirements
are discussed in details in the following sections.

The signal detection efficiency as well as the rate of pions, muons
from $\pi/K$ decays and muons from semi-muonic charmed hadron decays
entering the instrumented surface have been computed through a full
Monte Carlo simulation based on GEANT~3.21~\cite{geant3}. The target
is the LNGS rock, corresponding to a nearly isoscalar target with
density $\rho=2.71$~g/cm$^3$.  Neutrino interactions have been
produced by using the event generator described in Ref.~\cite{dario}
and final state particles are propagated through the rock by
accounting for all physical processes described in GEANT. Particles at
the exit of the rock are recorded and used for the analysis described
in the following.

\subsection{Signal}

\begin{figure}[tbhp]
\centering
\epsfig{file=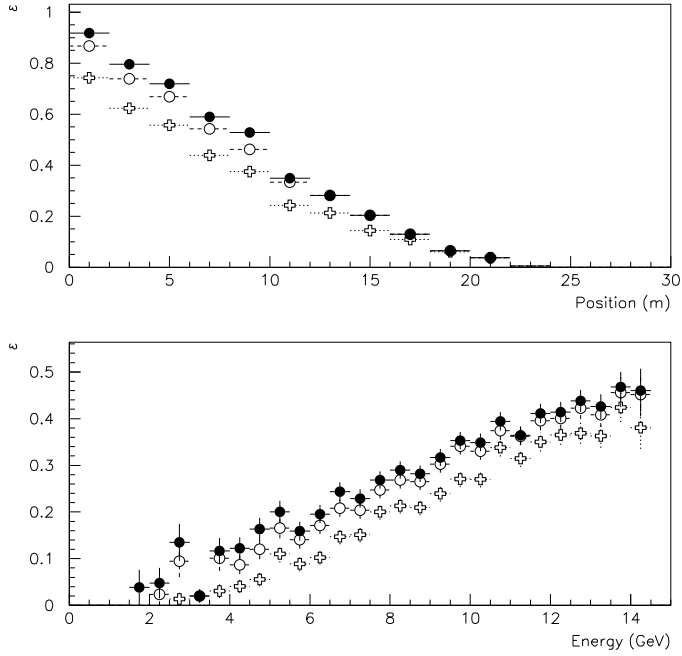,width=10cm}
\caption{ Probability for a muon (assuming 100\% $\nu_e\rightarrow
\nu_\mu$ oscillations) to exit from the rock as a function of the
interaction vertex (top panel) and of the neutrino energy (bottom
panel). The marks represent different energy cuts: larger than 0.5~GeV
(full circles), 1~GeV (empty circles) and 2~GeV (empty crosses). The
$\beta$-beam with $\gamma=2500$ is assumed.}
\label{fig:effenumucc_eff}
\end{figure}

\begin{figure}[tbhp]
\centering
\epsfig{file=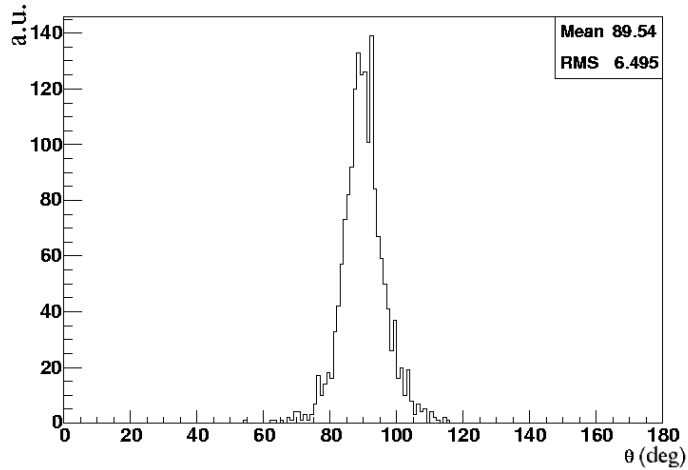,width=10cm}
\caption{Muon angular distribution, assuming 100\% $\nu_e\rightarrow
\nu_\mu$ oscillations, in $\nu_\mu$~CC interactions at the exit of the
rock for $E_\mu > 2$~GeV.}
\label{fig:thetamucut}
\end{figure}

\begin{figure}
\centering
\epsfig{file=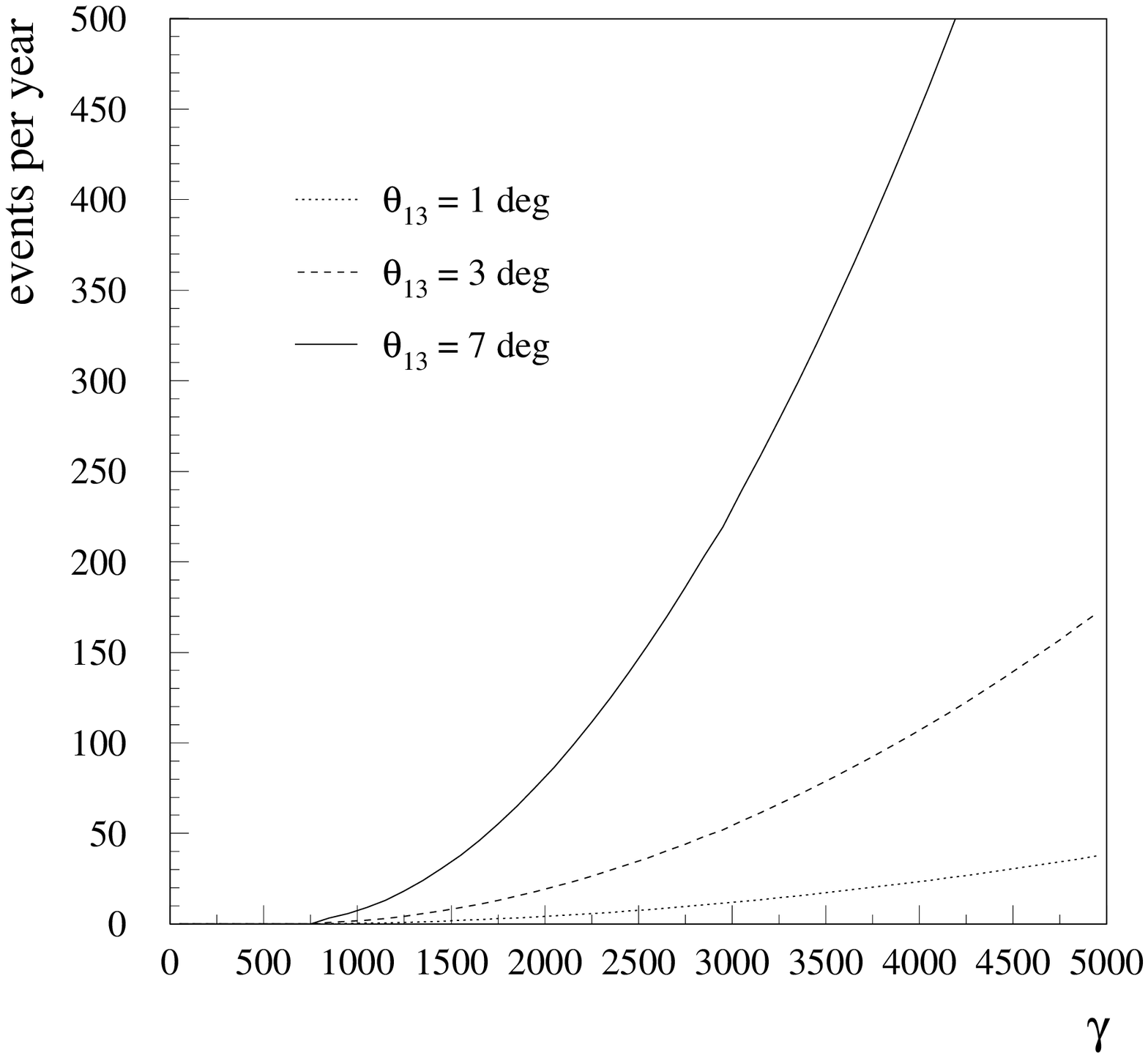,width=10cm}
\caption{$\nu_e \rightarrow \nu_\mu$ oscillated events per year versus
$\gamma$ for $\theta_{13}=1^\circ, 3^\circ,7^\circ$, $\delta=0^\circ$
and normal hierarchy. The rates include the detector efficiency
($E_\mu>2$~GeV) and are computed for one $15\times15 \ \mathrm{m}^2$
instrumented surface.}
\label{fig:osc_rate}
\end{figure}

An instrumented surface tags $\nu_e \rightarrow \nu_\mu$ oscillations
as an excess of horizontal multi-GeV muons plus a small $\nu_e
\rightarrow \nu_\tau \rightarrow \mu^- X$ contamination (``silver
channel''~\cite{silver}) that can be safely neglected here. The
direction of these muons corresponds approximately to the boost
direction of the CERN decay ring ($\sim 3^\circ$ below the horizon at
LNGS) smeared by multiple scattering in rock and iron. Their time
structure must be consistent with the time structure of the
circulating beams in the decay ring. A primary muon whose vertex is
located near the instrumented surface will reach the detector with
almost its original energy and in coincidence with other particles
belonging to the hadronic system. $\nu_\mu$~CC interactions occurring
deeper in the rock will have a cleaner topology due to the screening
of the accompanying hadrons but softer muons. Since the energy loss of
muons in rock is nearly linear with range ($\sim$2~MeV
g$^{-1}$cm$^2$), the target mass contributing to the overall event
rate grows linearly with the mean muon energy\footnote{Similar
considerations hold for $\bar{\nu}_\mu$ interactions.  However, at a
given (anti)neutrino energy, higher efficiencies than for $\nu_\mu$~CC
are expected due to the different $y\equiv 1-E_\mu/E_\nu$ dependence
of the cross section.}. Fig.~\ref{fig:effenumucc_eff} (top plot) shows
the probability for a muon to exit from the rock with an energy
greater than 0.5 (full circles), 1 (empty circles) and 2~GeV (empty
crosses) as a function of the interaction vertex. The muons come from
$\nu_\mu$~CC interactions in the rock and are computed assuming 100\%
$\nu_e \rightarrow \nu_\mu$ oscillations at $\gamma=2500$.  Their
angle distribution with respect to the horizontal direction is shown
in Fig.~\ref{fig:thetamucut} for $E_\mu>2$~GeV ($\theta=90^\circ$
corresponds to the boost direction of the decay ring). The muon
identification efficiency as a function of the parent neutrino energy
for a rock volume of 40$\times$15$\times$15~m$^3$ (24.4~kton) is shown
in Fig.~\ref{fig:effenumucc_eff} (bottom plot).  Finally, the number
of expected $\nu_e \rightarrow \nu_\mu$ events per year (corresponding
to $1.1\times 10^{18}~\neon$ decays) for $\theta_{13}=1^\circ,
3^\circ,7^\circ$, $\delta=0^\circ$ and normal hierarchy as a function
of $\gamma$ ($E_\mu>2$~GeV) is shown in Fig.~\ref{fig:osc_rate}.

\subsection{Beam related background}

The main sources of background having the same time structure of the
signal are the punch-through or decayed in flight (DIF) pions from the
bulk of $\nu_e$~CC and NC interactions in the rock and the semi-muonic
decay of charmed particles. Punch-through pions are mainly suppressed
by the instrumented iron acting as a pion plug (iron interaction
length $\lambda_I=16.76$~cm). The early decays in flight of pions
result into soft muons which are strongly reduced by the energy cut.
Table~\ref{table:background} (\ref{table:backgroundmu}) shows the
fraction of CC and NC interactions in a $40 \times 15 \times 15 \
\mathrm{m^3}$ rock volume giving a pion (muon) that enters the surface
with an energy greater than 0.5, 1, 2~GeV. These energy cuts
correspond to a range in iron of about 2.1,
4.2 and 8.4 interaction lengths. The first half of the table refers to
$\gamma=2500$ ($\nu_e$) and $\gamma=1500$ ($\bar{\nu}_e$); the second
one to $\gamma=4158$ ($\nu_e$) and $\gamma=2488$ ($\bar{\nu}_e$).

\begin{table}
\begin{center}
\begin{tabular}{|c|c|c|c|c|}
\hline
                              & No cut      & $p_{\pi^+(^-)}>0.5$~GeV &  $p_{\pi^+(^-)}>1$~GeV &  $p_{\pi^+(^-)}>2$~GeV \\
\hline
$\nu_e      $CC               & 1.00(0.62)\%&     0.70(0.38)\%        &   0.41(0.18)\%   & 0.17(0.07)\%      \\
\hline
$\nu_e      $NC               & 0.76(0.82)\%&     0.52(0.54)\%        &   0.27(0.29)\%   & 0.11(0.12)\%      \\
\hline
\hline
$\bar{\nu}_e$CC               & 0.18(0.41)\%&     0.10(0.25)\%        &   0.05(0.11)\%   & 0.018(0.024)\%      \\
\hline
$\bar{\nu}_e$NC               & 0.33(0.32)\%&     0.20(0.20)\%        &   0.09(0.09)\%   & 0.02(0.02)\%      \\
\hline
\hline
                              & No cut      & $p_{\pi^+(^-)}>0.5$~GeV &  $p_{\pi^+(^-)}>1$~GeV &  $p_{\pi^+(^-)}>2$~GeV \\
\hline
$\nu_e      $CC               & 1.59(1.07)\%&     1.01(0.71)\%        &   0.69(0.38)\%   & 0.35(0.19)\%      \\
\hline
$\nu_e      $NC               & 1.27(1.34)\%&     0.90(0.91)\%        &   0.53(0.53)\%   & 0.26(0.27)\%      \\
\hline
\hline
$\bar{\nu}_e$CC               & 0.36(0.69)\%&     0.24(0.46)\%        &   0.11(0.23)\%   & 0.04(0.08)\%      \\
\hline
$\bar{\nu}_e$NC               & 0.56(0.57)\%&     0.38(0.38)\%        &   0.20(0.19)\%   & 0.07(0.07)\%      \\
\hline
\end{tabular}
\caption{Probability for a $\bar{\nu}_e$, $\nu_e$ CC, NC event
generated in a $40 \times 15 \times 15 \ \mathrm{m^3}$ rock volume to
produce a primary or secondary pion that leaves the rock with a
momentum larger than a given cut. The first half of the table refers
to $\gamma=2500$ ($\nu_e$) and $\gamma=1500$ ($\bar{\nu}_e$); the
second half to $\gamma=4158$ ($\nu_e$) and $\gamma=2488$
($\bar{\nu}_e$).  The first number refers to $\pi^+$, the second
number (in parenthesis) to $\pi^-$.}
\label{table:background}
\end{center}
\end{table}

\begin{table}
\begin{center}
\begin{tabular}{|c|c|c|c|c|}
\hline
                              & No cut      & $p_{\mu^+(^-)}>0.5$~GeV &  $p_{\mu^+(^-)}>1$~GeV &  $p_{\mu^+(^-)}>2$~GeV \\
\hline
$\nu_e      $CC               & 0.030(0.012)\%&     0.016(0.004)\%        &   0.004(0.001)\%   & 0.001($<$0.001)\%      \\
\hline
$\nu_e      $NC               & 0.025(0.018)\%&     0.010(0.009)\%        &   0.002(0.004)\%   & 0.001(0.001)\%      \\
\hline
\hline
$\bar{\nu}_e$CC               & 0.007(0.011)\%&     0.003(0.005)\%        &   0.001(0.001)\%   & $<$0.001(($<$0.001)\%      \\
\hline
$\bar{\nu}_e$NC               & 0.011(0.008)\%&     0.004(0.002)\%        &   0.002(0.001)\%   & 0.001($<$0.001)\%      \\
\hline
\hline
                              & No cut      & $p_{\mu^+(^-)}>0.5$~GeV &  $p_{\mu^+(^-)}>1$~GeV &  $p_{\mu^+(^-)}>2$~GeV \\
\hline
$\nu_e      $CC               & 0.04(0.02)\%&     0.02(0.01)\%        &   0.008(0.006)\%   & 0.004(0.003)\%      \\
\hline
$\nu_e      $NC               & 0.03(0.03)\%&     0.015(0.014)\%        &   0.008(0.006)\%   & 0.002(0.001)\%      \\
\hline
\hline
$\bar{\nu}_e$CC               & 0.010(0.014)\%&     0.004(0.008)\%        &   0.001(0.003)\%   & 0.001(0.001)\%      \\
\hline
$\bar{\nu}_e$NC               & 0.018(0.012)\%&     0.007(0.004)\%        &   0.002(0.001)\%   & $<$0.001(0.001)\%      \\
\hline
\end{tabular}
\caption{ Probability for a $\bar{\nu}_e$, $\nu_e$ CC, NC event
generated in a $40 \times 15 \times 15 \ \mathrm{m^3}$ rock volume to
produce a muon that leaves the rock with a momentum larger than a
given cut. The first half of the table refers to $\gamma=2500$
($\nu_e$) and $\gamma=1500$ ($\bar{\nu}_e$); the second half to
$\gamma=4158$ ($\nu_e$) and $\gamma=2488$ ($\bar{\nu}_e$). The first
number refers to $\mu^+$, the second number (in parenthesis) to
$\mu^-$.}
\label{table:backgroundmu}
\end{center}
\end{table}

The charm production rate has been estimated by using the latest
results from the CHORUS~\cite{chorus} experiment and accounting
for the different (anti)neutrino energy spectrum. For $\gamma$ of the
order of $2500(\nu_e)/1500(\bar{\nu}_e)$ the cross sections are 2.3\%
and 1.8\% for $\nu_e$ and $\bar{\nu}_e$, respectively. They grow up to
3.5\% ($\nu_e$) and 2.7\% ($\bar{\nu}_e$) for
$\gamma=4158(\nu_e)/2488(\bar{\nu}_e)$.
The semi-muonic branching ratio is about 7\%.  Note that in the highest
$\gamma$ scenario, the muon spectrum from charm is significantly
harder than in the region close to the kinematic threshold and the
contamination increases substantially. This effect is described in
Table~\ref{charm-back} where the expected background 
at $\gamma=2500(\nu_e)/1500(\bar{\nu}_e)$ and
$\gamma=4158(\nu_e)/2488(\bar{\nu}_e)$ with different momentum cuts
are shown.  It is worth noting that this background would be
substantially reduced if the sign of the muon were available, its
charge being opposite with respect to the primary $\mu$ from
$\nu_\mu$~CC. For such a soft muon spectrum, a large-surface
magnetized iron detector with a field of about 1~Tesla would have a
rejection factor greater than 99\% and, in addition, would suppress
the punch-through and DIF background by about 50\%.

\begin{table}
\begin{center}
\begin{tabular}{|c|c|c|c|c|}
\hline
 & No cut & $p_\mu>0.5$~GeV &  $p_\mu>1$~GeV  &  $p_\mu>2$~GeV  \\
\hline
$\nu_e      $CC ($\gamma=2500$)  &   6.46 \%       &  6.33  \%         &  5.85 \%           & 2.92 \%          \\
\hline
$\bar{\nu}_e$CC  ($\gamma=1500$) &    5.51 \%        &  5.51 \%          &   3.94 \%           &   0.95 \%          \\
\hline
\hline
  & No cut & $p_\mu>0.5$~GeV &  $p_\mu>1$~GeV  &  $p_\mu>2$~GeV  \\
\hline
$\nu_e      $CC ($\gamma=4158$) &  10.63 \%        & 10.53 \%          & 10.31 \%           & 7.63 \%           \\
\hline
$\bar{\nu}_e$CC  ($\gamma=2488$) &  7.15 \%         & 6.89 \%           & 5.92 \%            & 3.44 \%           \\
\hline
\end{tabular}
\caption{Probability for a semi-muonic decay of charmed $\nu_e$
($\bar{\nu}_e$) CC event generated in a $40 \times 15 \times 15 \
\mathrm{m^3}$ rock volume to produce a muon that leaves the rock with
a momentum larger than a given cut. The first half of the table refers
to $\gamma=2500$ ($\nu_e$) and $\gamma=1500$ ($\bar{\nu}_e$); the
second half to $\gamma=4158$ ($\nu_e$) and $\gamma=2488$
($\bar{\nu}_e$).}
\label{charm-back}
\end{center}
\end{table}

\subsection{Beam unrelated background}

\begin{figure}
\centering
\epsfig{file=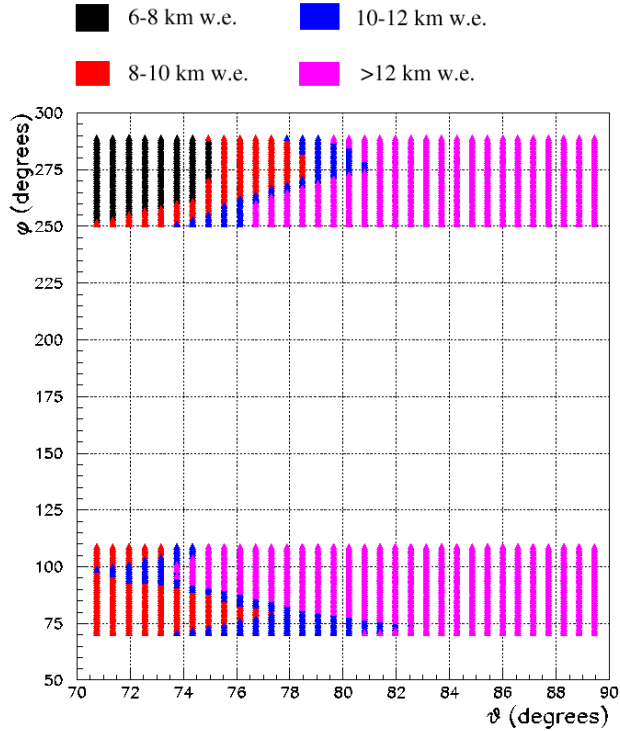,width=10cm}
\caption{Rock coverage expressed in km of water equivalent (km w.e.)
in the range $\theta \simle 20^\circ$ with respect to the beam
direction as a function of the zenith and azimuth angle (courtesy of
the LVD collaboration).  The direction pointing to CERN corresponds to
$\theta \sim 90^\circ$ and $\phi \sim 90^\circ$ (see text for
details).}
\label{fig:rock_coverage}
\end{figure}

The LNGS underground halls are located at a slant depth that strongly
depends on the zenithal and azimuthal direction. The minimum depth is
$\sim$3200~hg/cm$^2$ but, due to a fortunate conspiracy of the mountain
profile, the largest rock depths near the horizon are reached in the
CERN-to-LNGS direction. The rock coverage in the zenith ($\theta$) and
azimuth ($\phi$) region pointing to CERN is shown in
Fig.~\ref{fig:rock_coverage}. Here, $\theta=90^\circ$ corresponds to
the horizon; at $\phi \sim 90^\circ$ the muons come from the CERN
direction and enter the instrumented surface from the front. At $\phi
\sim 270^\circ$ muons enter the detector from the back.  The latter
can be vetoed if the active detectors provide proper timing.  As can
be inferred from Fig.~\ref{fig:rock_coverage}, the slant depth is
greater than 12 km of water-equivalent (km w.e.) in most of the region
of interest. Here, the muon flux is dominated by atmospheric neutrinos
and it is of the order of $\sim 5 \times 10^{-13}\ \mathrm{cm}^{-2} \
\mathrm{s}^{-1} \mathrm{sr}^{-1}$~\cite{macro_atm}.
A detailed calculation has been carried out through a full
parametrization of the data from the MACRO
experiment~\cite{MACRO_muons}. Results are shown in
Table~\ref{table:cosmback}.  It is clear that, even at the largest
angle ($\theta \sim 40^\circ$), the time structure of the beam
($10^{-4}$ suppression factor) allows a complete rejection of the
background.  In fact, this result suggests that, opposite to the
low-$\gamma$ option (CERN to Frejus), the constraint on the bunch
length of the beta beam ($<14$~ns) can be released by at least one
order of magnitude.

\begin{table}
\begin{center}
\begin{tabular}{|c|c|c|c|c|c|c|c|}
\hline
        & All    & $p_\mu>0.5 ~\mathrm{GeV} $ &  $\mu_{\mathrm{fb}}$& $\mu_{\mathrm{f}}$ &  $\theta_\mu < 20^\circ$ &  $\theta_\mu < 30^\circ$ &  $\theta_\mu < 40^\circ$ \\
\hline
$\mu^-$ & 425945 & 424537             & 301795                 & 100681             & 19                      & 126                     & 728                     \\
$\mu^+$ & 510743 & 509033             & 361332                 & 120797             & 27                      & 177                     & 928                     \\
\hline
\hline
        & All    & $p_\mu>1 ~\mathrm{GeV} $ &  $\mu_{\mathrm{fb}}$& $\mu_{\mathrm{f}}$ &  $\theta_\mu < 20^\circ$ &  $\theta_\mu < 30^\circ$ &  $\theta_\mu < 40^\circ$ \\
\hline
$\mu^-$ & 425945 & 423105             & 300753                 & 100362             & 19                      & 126                     & 727                     \\
$\mu^+$ & 510743 & 507273             & 360088                 & 120391             & 27                      & 176                     & 922                     \\
\hline
\hline
        & All    & $p_\mu>2 ~\mathrm{GeV} $ &  $\mu_{\mathrm{fb}}$& $\mu_{\mathrm{f}}$ &  $\theta_\mu < 20^\circ$ &  $\theta_\mu < 30^\circ$ &  $\theta_\mu < 40^\circ$ \\
\hline
$\mu^-$ & 425945 & 420298             & 298738                 & 99703              & 19                      & 125                     & 725                     \\
$\mu^+$ & 510743 & 503765             & 357664                 & 119578             & 27                      & 173                     & 918                     \\
\hline
\end{tabular}
\caption{Number of underground muons surviving different sets of cuts
in 1 year data taking with one basic unit target. $\mu_{\mathrm{fb}}$
stands for the number of cosmic muons entering the detector from the
front and back side; $\mu_{\mathrm{f}}$ includes just the muons from
the front. The suppression factor due to the time structure of the
beam is not taken into account.}
\label{table:cosmback}
\end{center}
\end{table}

\subsection{Summary of expected rates}
\label{sec:exprates}

From the above discussion, it is apparent that the sensitivity of the
detector under consideration will be limited mainly by the beam
related background. A severe requirement on the visible range of the
penetrating tracks, corresponding to an energy cut of about 2~GeV,
will bring the punch-trough contamination at the level of a few events
per year (suppression factor $< 10^{-3}$). The charm contamination is
expected to limit the sensitivity at the highest $\gamma$ (see
Table~\ref{charm-back}); a further suppression factor of the charm
background ($< 10^{-2}$) from charge reconstruction is available for a
magnetized detector\footnote{In the few GeV energy range, a magnetized
iron detector with $B\simeq 1 \ \mathrm{T}$ and, if necessary,
precision trackers before and after the iron plug can achieve charge
misidentification probabilities well below $10^{-2}$~\cite{silver}. In
order to determine the actual efficiency, an optimization of the
detector and of the pattern recognition algorithm is
mandatory. Clearly, this issue is beyond the scope of this paper; in
the following we assume very conservatively a 99\% charge
identification efficiency.}. In most of the cases, however, the
sensitivity is limited by the contamination of secondary muons from
$\pi$ and $K$ decay in flight. A synopsis of the expected rates per
year for a $15 \times 15 \ \mathrm{m}^2$ instrumented surface in the
occurrence of the null hypothesis ($\theta_{13}=0^\circ \ \
\Longrightarrow P(\nu_e \rightarrow \nu_\mu) \simeq O_4$ ) for $E_\mu
>2 \ \mathrm {GeV}$ and an angle $\theta<40^\circ$ with respect to the
nominal beam direction
is shown in Table~\ref{table:events}. The events per year expected for
100\% $\nu_e \rightarrow \nu_\mu$ conversion probability are $9.3
\times 10^{4}$ ($\nu_e$ at $\gamma=2500$), $2.0 \times 10^{4}$
($\bar{\nu}_e$ at $\gamma=1500$), $7.9 \times 10^{5}$ ($\nu_e$ at
$\gamma=4158$) and $2.1 \times 10^{5}$ ($\bar{\nu}_e$ at
$\gamma=2488$).

\begin{table}
\begin{center}
\begin{tabular}{|c|c|c|c|c|c|}
\hline
Detector & $\gamma$ & $\nu_e \rightarrow \nu_\mu$ & $\pi$ & $\mu$ & charm \\ 
\hline
$B=0$~T   &  $2500 \ (\nu_e)$        &   1.5  & 0.5 &  11.6  & 20.2   \\   
$B=0$~T  &  $1500 \ (\bar{\nu}_e)$  &   0.8  & 0.02 &  3.5  & 1.5  \\
$B=0$~T   &  $4158 \ (\nu_e)$        &   4.9  & 4.6 &  153.4 & 357.1  \\
$B=0$~T   &  $2488 \ (\bar{\nu}_e)$  &   3.2  & 0.3 &  15.4  & 37.1   \\

\hline \hline

$B\sim 1$~T  &  $2500 \ (\nu_e)$        & 1.5 & 0.2  & 5.8 & 0.2   \\
$B\sim 1$~T  &  $1500 \ (\bar{\nu}_e)$  & 0.8 & 0.01 & 1.8  & 0.01  \\
$B\sim 1$~T  &  $4158 \ (\nu_e)$        & 4.9 & 1.8  & 64.8 & 3.6   \\
$B\sim 1$~T  &  $2488 \ (\bar{\nu}_e)$  & 3.2 & 0.1  & 7.8 & 0.4   \\
\hline
\end{tabular}
\caption{Number of event per year for a $15 \times 15 \ \mathrm{m}^2$
instrumented surface in the occurrence of the null hypothesis
($\theta_{13}=0^\circ$). $B\sim 1$~T ($B=0$) refers to the detector option with
(without) magnetic field.}
\label{table:events}
\end{center}
\end{table}

\section{Sensitivity}
\label{sec:sensitivity}

\begin{figure}
\centering
\epsfig{file=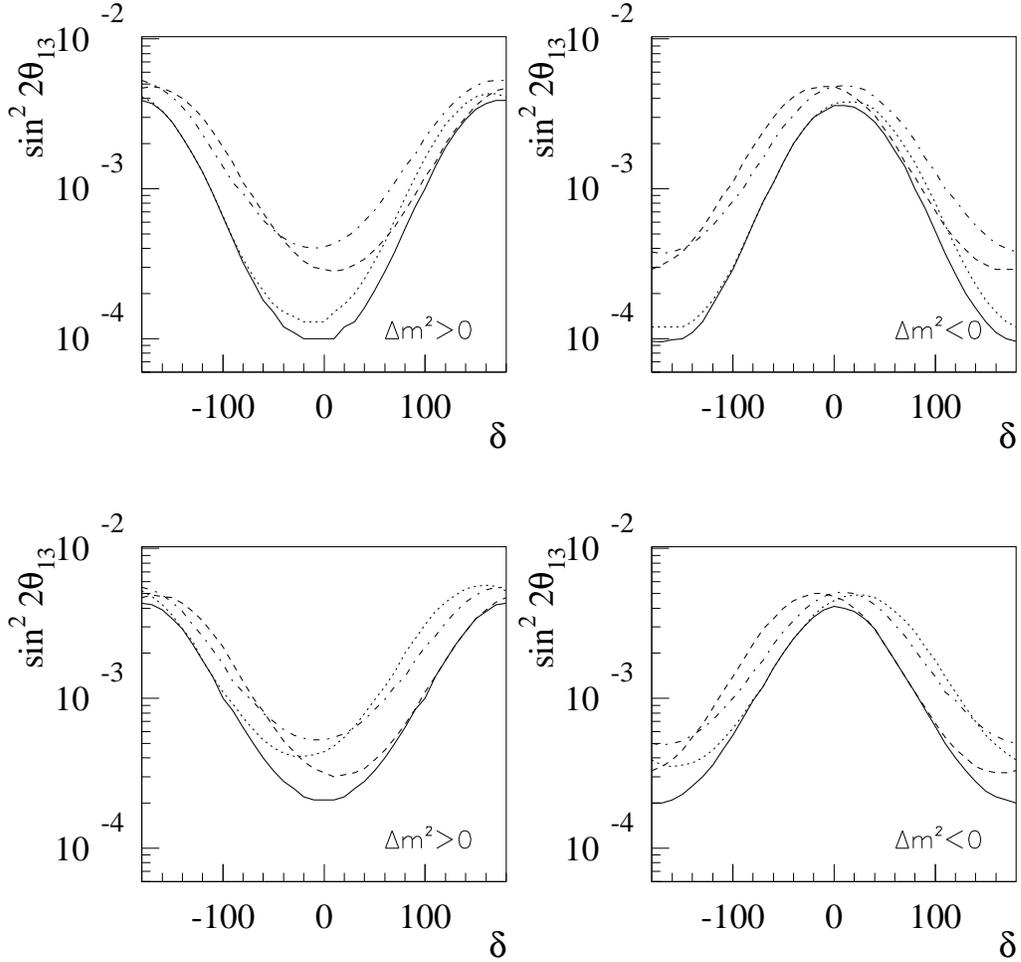,width=\textwidth}
\caption{ 90\% C.L. excluded region in the occurrence of the null
hypothesis ($\theta_{13}=0^\circ$) for normal (left plots) and
inverted hierarchy (right plots) and for
$\gamma=2500(\nu_e)/1500(\bar{\nu}_e)$ (lower plots) and
$\gamma=4158(\nu_e)/2488(\bar{\nu}_e)$ (upper plots) using the
neutrinos coming from $\neon$ decays (dots), the antineutrinos from
$\helio$ (dashed line) and combining the two measurements (solid
line). The dot-dashed line represents the sensitivity ($\nu$ and
$\bar{\nu}$ combined) for a detector without magnetic field.}
\label{fig:null_hyp}
\end{figure}

\begin{figure}
\centering
\epsfig{file=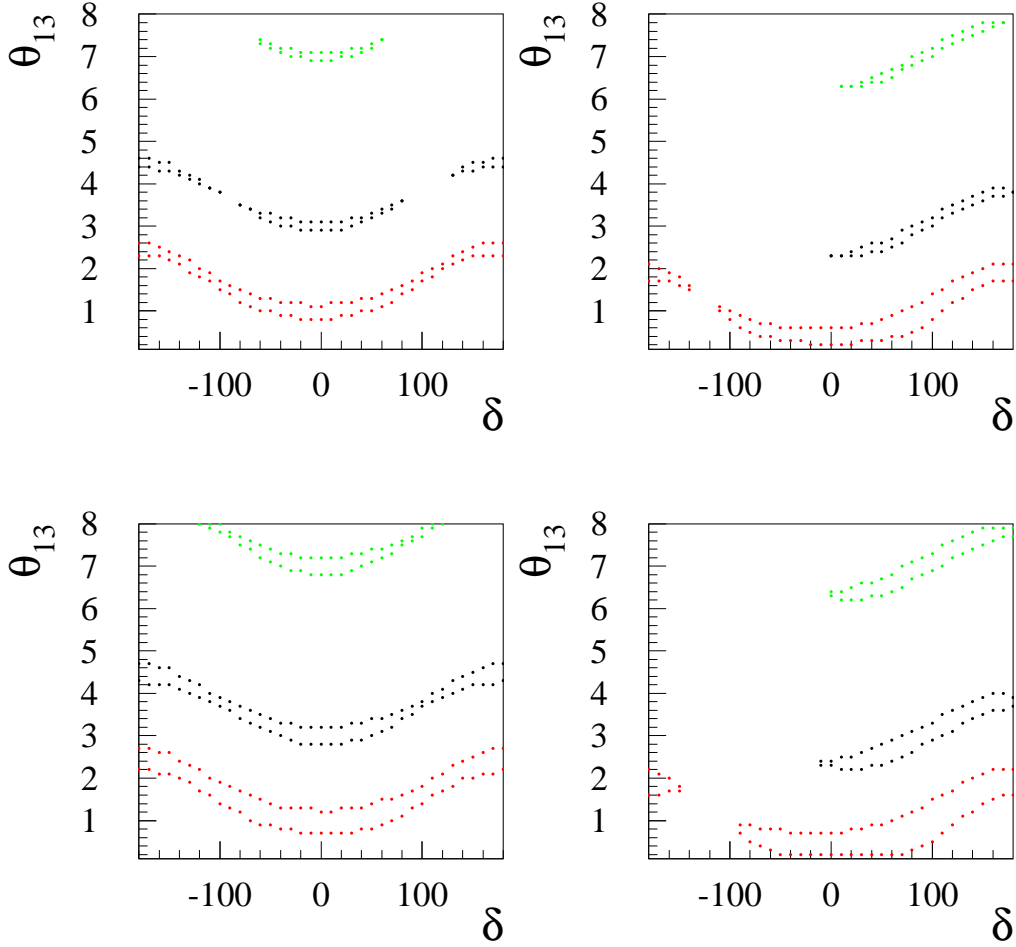,width=\textwidth}
\caption{ Allowed regions at 90\% C.L. for $\theta_{13}=1^\circ$
 (red), $3^\circ$ (black), $7^\circ$ (green), $\delta=0^\circ$ (left
 plots) and $\delta=90^\circ$ (right plots) for normal hierarchy and a
 magnetized detector operating at
 $\gamma=2500(\nu_e)/1500(\bar{\nu}_e)$ (lower plots) and
 $\gamma=4158(\nu_e)/2488(\bar{\nu}_e)$ (upper plots).  Both the
 neutrino and the antineutrino fluxes are combined.}
\label{fig:positive_result}
\end{figure}

\begin{figure}
\centering
\epsfig{file=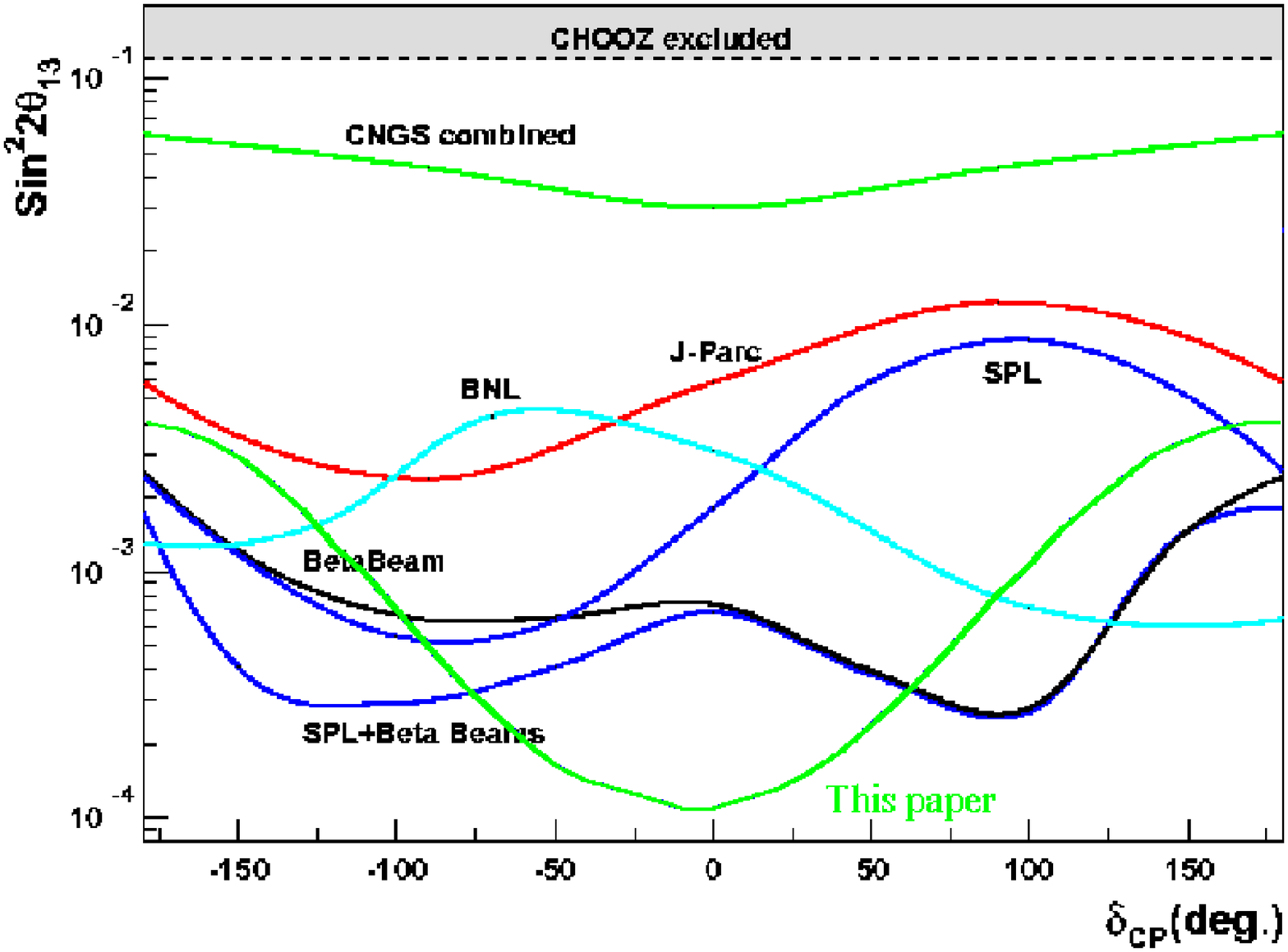,width=12cm}
\caption{ 90\% C.L. excluded region in the occurrence of the null
hypothesis ($\theta_{13}=0^\circ$) for normal hierarchy
and various experimental facilities~\cite{Bouchez:2003fy,comparison}
(see text for details).}
\label{fig:comparison}
\end{figure}

The sensitivity to the $\theta_{13}$ and $\delta$ parameters is
evaluated assuming an instrumented surface of $15\times15$~m$^2$ with
a detector having an iron depth greater than 8 interaction
lengths. Only muons with energy greater than 2~GeV at the entrance of
the detector are considered. A 5 year data taking with $2.9\times
10^{18}~\helio$ and $1.1\times 10^{18}~\neon$ decays per year is
assumed\footnote{Note that, in current literature, sometimes results
for the low-$\gamma$ CERN to Frejus option are given assuming 10 years
of data taking.}. The baseline $L$ corresponds to the CERN to Gran
Sasso distance (732~km). Results are provided for two high-$\gamma$
options: $\gamma=2500(\nu_e)/1500(\bar{\nu}_e)$ and
$\gamma=4158(\nu_e)/2488(\bar{\nu}_e)$. The former is dominated by the
pion background from $\nu_e$~CC and NC. The latter suffers from a
significant charm contamination which can be eliminated by charge
reconstruction (magnetized iron detector option). The sensitivity has
been computed fixing all parameters but $\theta_{13}$ and $\delta$ to
their current best values ($\Delta m^2_{21}=7.3 \times
10^{-5}$~eV$^2$, $\sin^2 2\theta_{12}=0.8$, $|\Delta m^2_{31}|=2.5
\times 10^{-3}$~eV$^2$, $\sin^2 2\theta_{23}=1$) and performing a
two-parameter $\chi^2$ fit. The facility is used as a pure counting
experiment and the information coming from the reconstructed muon
spectrum at the entrance of the instrumented surface is not
exploited. Fig.~\ref{fig:null_hyp} shows the parameter region excluded
at 90\% C.L. in the occurrence of the null hypothesis
($\theta_{13}=0^\circ$) for normal (left plots) and inverted hierarchy
(right plots) and for $\gamma=2500(\nu_e)/1500(\bar{\nu}_e)$ (lower
plots) and $\gamma=4158(\nu_e)/2488(\bar{\nu}_e)$ (upper plots). Here,
we assume reconstruction of the muon charge, use the neutrinos coming
from $\neon$ decays (dots), the antineutrinos from $\helio$ (dashed
line) and combine the two measurements (solid line). The dot-dashed
line represents the sensitivity for a detector without magnetic field.
Note that, in spite of the reduced cross section, the antineutrino
flux strongly contributes to the sensitivity of the apparatus due to
the lower background contamination. As noted above, the suppression of
matter effect does not allow an unique determination of the neutrino
hierarchy so that $\mathrm{sign} \ \left[\delta \cdot
\mathrm{sign}(\Delta m^2_{13} ) \right] $ remains undetected.  The
allowed regions at 90\% C.L. for $\theta_{13}=1^\circ$ (red),
$3^\circ$ (black), $7^\circ$ (green), $\delta=0^\circ$ (left plot) and
$\delta=90^\circ$ (right plot) are shown in
Fig.~\ref{fig:positive_result} (lower plots) for normal hierarchy, a
magnetized detector operating at
$\gamma=2500(\nu_e)/1500(\bar{\nu}_e)$ and combining both the neutrino
and the antineutrino flux.  The corresponding curves for
$\gamma=4158(\nu_e)/2488(\bar{\nu}_e)$ are shown in
Fig.~\ref{fig:positive_result} (upper plots).  The sensitivity of this
apparatus compared with other proposed
facilities~\cite{Bouchez:2003fy,comparison} is shown in
Fig.~\ref{fig:comparison}. The facilities under consideration are the
CERN to Gran Sasso $\nu$ beam (CNGS), JAERI to Kamioka (J-Parc), BNL
to Homestake (BNL), the low-$\gamma$ beta beam (``Beta beam'') and the
SPL to Frejus superbeam (SPL). The current CHOOZ limit is also shown.
The label ``This paper'' refers to the
$\gamma=4158(\nu_e)/2488(\bar{\nu}_e)$ magnetized detector
configuration after 5 years of data taking.

\section{Conclusions}
The $\nu$ fluxes at large distance produced by $\beta$ decays of
boosted radioactive ions has a strong quadratic dependence on the
Lorentz factor $\gamma$ of the ions. For a fixed baseline, an increase
of $\gamma$ that brings the average neutrino energy well above the
maximum of the oscillation probability does not imply a loss of events
since the increase of the flux compensates for the smallness of the
oscillation probability. Moreover, a net gain of events is obtained by
the linear rise of the CC cross section and by the increase of the
detector efficiency. The latter effect is particularly valuable for
purely passive detectors.  In this paper we propose to observe $\nu_e
\rightarrow \nu_\mu$ oscillations as an excess of horizontal muons
produced in the rock and tracked by a low-mass instrumented surface
installed in an underground hall of LNGS. This configuration allows a
very precise determination of $\theta_{13}$ and turns out to be
competitive with the current low-$\gamma$ design based on a gigantic
water Cherenkov at Frejus~\cite{mton_at_frejus}.  Similarly to the
latter, it has limited sensitivity to the sign of $\Delta
m^2_{13}$. However, opposite to the facilities operating at the peak
of the oscillation maximum, it shows maximal $\theta_{13}$ sensitivity
for small CP violation ($\delta \simeq 0, \pm \pi$).  For
$\delta=\pi/2$, this complex would significantly constrain the
$\theta_{13},\delta$ parameter space (see
Fig.~\ref{fig:positive_result}) but, in general, would not tag
explicitly CP violation through a lepton/antilepton asymmetry, since
the off-peak configuration suppresses the CP odd terms. Clearly, the
cost of the detector is negligible compared e.g. with a Mton water
Cherenkov, while most of the acceleration system is either shared with
other non-neutrino projects (EURISOL) or based on existing CERN
machines.

\section*{Acknowledgments}
We wish express our gratitude to A.~Marini, F.~Ronga and L.~Votano for
many useful suggestions. We thank M.~Selvi and M.~Sioli for providing
us with current data on cosmic background at LNGS. A special thank to
M.~Mezzetto for interesting comments and discussions concerning the
physics and technological challenges of the beta beams.

%


\begin{thebibliography}{999}
%

\bibitem{Zucchelli:sa}
P.~Zucchelli, Phys.\ Lett.\ B {\bf 532} (2002) 166.

\bibitem{PMNS}
B.~Pontecorvo, Sov.\ Phys.\ JETP {\bf 6} (1957) 429;
Z.~`<Maki, M.~Nakagawa, S.~Sakata, Prog.~Theor.~Phys.~{\bf 28}~(1962)~870.

\bibitem{Mezzetto:2003ub}
M.~Mezzetto, J.\ Phys.\ G {\bf 29} (2003) 1781.

\bibitem{Bouchez:2003fy} J.~Bouchez, M.~Lindroos and M.~Mezzetto,
Contributed to 5th International Workshop on Neutrino Factories and
Superbeams (NuFact 03), New York, 5-11 Jun 2003;
arXiv:hep-ex/0310059.

\bibitem{Burguet-Castell:2003vv} 
J.~Burguet-Castell, D.~Casper,
J.~J.~Gomez-Cadenas, P.~Hernandez and F.~Sanchez,
Nucl. Phys. B {\bf 695} (2004) 217 [arXiv:hep-ph/0312068].

\bibitem{betabeams_moriond} Workshop on ``Radioactive beams
for nuclear physics and neutrino physics'' 37$^{th}$ Rencontre de
Moriond, Les Arcs (France) March 17-22nd, 2003; 
http://moriond.in2p3.fr/radio/index.html.

\bibitem{mton_at_frejus}
L. Mosca, ``A European Megaton Project at Frejus'',
Talk at the 8$^{th}$ International Workshop on
Topics in Astroparticle and Underground Physics (TAUP2003),
Seattle, Washington, September 2003.

\bibitem{Migliozzi:2003pw}
P.~Migliozzi, F.~Terranova,
Phys.\ Lett.\ B {\bf 563} (2003) 73.

\bibitem{Autin:2002ms}
B.~Autin {\it et al.},
J.\ Phys.\ G {\bf 29} (2003) 1785.

\bibitem{SPL} B. Autin {\it et al.},''Conceptual design of the SPL, a
high-power superconducting $H^-$ linac at CERN'', CERN-2000-012.

\bibitem{Apollonio:2002en}
M.~Apollonio {\it et al.}, CERN-TH-2002-208,
arXiv:hep-ph/0210192.

\bibitem{talk_lindroos}
M.~Lindroos, ``Accelerator based neutrino beams'', 
in \cite{betabeams_moriond}.

\bibitem{cervera_freund}
A.~Cervera {\it et al.}, Nucl. Phys. {\bf B579}~(2000)~17, erratum ibid.
Nucl. Phys. {\bf B593}~(2001)~731; 
M.~Freund, Phys. Rev.~{\bf D64}~(2001)~053003.

\bibitem{Maltoni:2003da}
M.~Maltoni, T.~Schwetz, M.~A.~Tortola, J.~W.~F.~Valle,
Phys.\ Rev.\ D {\bf 68} (2003) 113010.

\bibitem{geant3} GEANT Detector Description and Simulation Tool, CERN
Program Library long Writeup W5013, Geneve, 1993.

\bibitem{dario}
D.~Autiero, ``The OPERA event generator and the data tuning of nuclear 
reinteractions'', Talk at the $3^{rd}$ International Workshop on
Neutrino-Nucleus Interactions in the Few GeV Region (NuInt04),
March 17-21, 2004, Assegi, Italy.

\bibitem{silver}
A.~Donini, D.~Meloni and P.~Migliozzi,
Nucl.\ Phys.\ B {\bf 646} (2002) 321; 
D.~Autiero {\it et al.},
Eur.\ Phys.\ J.\ C {\bf 33} (2004) 243.

\bibitem{chorus} 
A.~Artamonov, ``Cross section measurements in CHORUS'', Talk at the
$3^{rd}$ International Workshop on Neutrino-Nucleus Interactions in
the Few GeV Region (NuInt04), March 17-21, 2004, Assegi, Italy.

\bibitem{macro_atm}
M.~Ambrosio {\it et al.}  [MACRO Coll.], Phys.\ Lett.\ B {\bf 517}
(2001) 59.

\bibitem{MACRO_muons}
Courtesy of M.~Sioli.

\bibitem{comparison} The plot is based on the analyses of M.~Apollonio
{\it et al.}, Phys. \ Lett. \ B {\bf 466} (1999) 415; Y.~Itow {\em et
al.}, arXiv:hep-ex/0106019; M.~V.~Diwan {\it et al.}, Phys.\ Rev.\ D
{\bf 68}, 012002 (2003); M. Mezzetto, J. \ Phys. \ G {\bf 29} (2003)
1771 and of Refs.\cite{Mezzetto:2003ub,Migliozzi:2003pw}


\end{thebibliography}
\end{document}